\documentclass[12pt]{article}
\usepackage{qbnets_dmats}
\begin{document}
\title{An Introduction to\\
Quantum Bayesian Networks for \\
Mixed States}

\author{Robert R. Tucci\\
        P.O. Box 226\\
        Bedford,  MA   01730\\
        tucci@ar-tiste.com}

\date{\today}
\maketitle
\vskip2cm
\section*{Abstract}
This paper is intended to be
a pedagogical introduction to
quantum Bayesian networks (QB nets),
as I personally use them
to represent mixed states (i.e., density matrices,
and open quantum systems).
A special
effort is made to make
contact with notions
 used in textbooks on quantum Shannon
Information Theory (quantum SIT),
such as the one
by Mark Wilde (arXiv:1106.1445)

\newpage

\section{Introduction}
This paper is intended to be
a pedagogical introduction to
quantum Bayesian networks (QB nets),
as I personally use them
to represent mixed states (i.e., density matrices,
and open quantum systems).
A special
effort is made to make
contact with notions
 used in textbooks on quantum Shannon
Information Theory (quantum SIT),
such as the one
by Mark Wilde\cite{Wilde}.

QB nets are a generalization
of classical Bayesian networks (CB nets)
to quantum mechanics.
CB nets have been hot topic
in AI circles since
the seminal work of Judea Pearl
and collaborators
that started in the 1980's.
A very complete book on CB nets
is the one by Koller and Friedman\cite{KF}.

Just like mankind has devised
many names for the idea of God,
there are other names for CB nets
and variations of the idea.
Others have called them causal
probabilistic diagrams, factor
graphs, probabilistic
system diagrams, etc. To be sure,
there are some differences between
some of these diagrams and CB nets, but
all seem to be striving to
conjure up the same divine concept.
Some of the close siblings of
CB nets are discussed in Ref.\cite{Loe},
an IEEE magazine article
by Loeliger.

One variation
on the CB net idea
involves using graphs
in which the arrows (a.k.a.
directed edges) represent
tensor indices and the
boxes (a.k.a. nodes, vertices) represent
transition matrices. In
this approach, call it
the tensor-graphs approach, each arrow
coming out of a fixed node
carries different stuff.
In the CB nets approach, the nodes
again represent transition
matrices, but the arrows
perform a very different job.
Each arrow
coming out of a fixed node
carries the same stuff,
namely the name
of the node the arrow originates at.
This difference might
appear subtle or even insignificant to
the untrained eye, but Bayesian network
believers (like me) swear by it,
claiming that it is clearer and more
powerful than the tensor-graphs approach
{\it when
dealing with probabilities}.
Bayesian network believers think
that using tensor graphs
to describe probability networks
is like trying to fill round holes
with square pegs. Even when the pegs
fit, they don't do a very good job.

Classical information theorists
have been using tensor-graphs
in their field for a long time.
See, for example, the chapter
on ``network information theory" in
the book by Cover and Thomas, Ref.\cite{CovTh}.
Or look at the book
by El Gamal and Kim, Ref.\cite{ElG}, which
is devoted exclusively
to the subject of network information theory.

Quantum information theorists have
been using tensor-graphs
 in their field
for a long time too, at least
since the seminal work by Schumacher
and collaborators. For an early example
of a paper by
Schumacher that uses
tensor-graphs, see, for example, Ref.\cite{Schum},
published in 1996.

Some early quantum information papers,
for example the seminal paper
Ref.\cite{Ben} by Bennett et al.
also use a version
of tensor-graphs,
but they use them
in a very loose, ambiguous,
imprecise way.
Ref.\cite{Ben} even has
some diagrams
that sound like
 sacrilege to the ears of this Bayesian nets
believer, such as diagrams
that have nodes intended to
represent buckets of sewage
(Fig.13 in Ref.\cite{Ben}).

Not only do adherents to ``the QB net way"
espouse being
good parents by treating all arrows (= children)
coming
out of a fixed node (= parent)
the same. We are also very
strict about our nodes. Each node
has a numerical ``value",
and the whole graph also has a value
which equals the product
of the node values.
Nodes aren't there
just for decoration, or as mere labels
with no numerical value assigned to them,
or only to convey an abstract
notion like ``do this operation now".

Many quantum information papers
don't use diagrams at all. They
specify their quantum ``protocols" or
algorithms in terms of ``pseudo
code". In my opinion,
those papers would be much clearer
if they described their
algorithms using both, pseudo-code and
QB nets, whenever this is possible.
The recent book on quantum information
theory by Mark Wilde\cite{Wilde}
earns high marks in this regard, as it
uses a lot of diagrams.
Wilde's diagrams also have a fairly
precise meaning. However,
they are tensor-graphs, not
the wonderful QB nets.

My own
work on QB nets started about 15 years ago
with
Ref.\cite{Tuc97}.
In that early paper, I
dealt only with QB nets for pure quantum
states.
I've been using QB nets for mixed states
since at least Ref.\cite{Tuc07}.
This paper repeats some of
the ideas of Ref.\cite{Tuc07}
for QB nets of mixed states,
with (hopefully) some
small improvements.

I've also written a Mac application
that does QB nets
called Quantum Fog.\footnote{
I stopped Quantum Fog development in 2006.
The application is still available for free
at Ref.\cite{QFog}.
Quantum Fog's last
version (Version 2.0) is known to work
with Mac OS X $\leq$ 10.4.
It probably works with some
higher versions
of Mac OS X too.
 Quantum Fog only
does QB nets for pure states.
That's not
because QB nets can't deal
with mixed states as some
people think. It's only because
I stopped developing Quantum Fog before
I had a chance to add to it the
capability
to do mixed state calculations.}

I also have a blog called
Quantum Bayesian Networks (Ref.\cite{qbnets})
in which I regularly post
articles about Bayesian networks and quantum
computing.

Subsequent to Refs.\cite{Tuc97, Tuc07},
other workers have
devised their own types of diagrams
for doing quantum information theory.
Their diagrams are very different
from the QB nets in
this paper.

\begin{enumerate}
\item[(a)]
Leifer and Poulin in Ref.\cite{LP},
and later Leifer and Spekkens in Ref.\cite{LS},
postulate some directed acyclic graphs,
but {\it they assign a whole density
matrix to each node of the graph}.
Furthermore, their node
density matrices would be quite hard
to calculate in practice, especially for
complicated graphs.
In comparison, a whole QB net is
used to describe one density matrix.
And the transition matrix
that a QB net assigns to each
node comes from the definition
of a probability amplitude,
a really basic thing that requires
almost no calculation---
certainly less calculation
than the node density matrices of Leifer
and coworkers.

I like to think of  QB nets as: A
light container of data,
useful as a data structure in
computer programming.
A vehicle rather than a destination.
A transparent, tidy way
of organizing a lot of data
in a pictorial way, {\it prior} to intensive
calculation, not as
being itself
the outcome of a major calculation.

\item[(b)]
Coecke (Ref.\cite{Coe}) and collaborators
use
category theory
to define their diagrams. In
comparison, QB nets
are much less abstract.
Defining them requires no category theory,
just standard, run-of-the-mill
quantum mechanics.
\end{enumerate}

The QB nets in this paper
are much
simpler than
the diagrams of
(a) and (b).
Simplicity
can be a virtue
in mathematics (consider, for example,
abstract algebra's
definition of a group,
which is simplicity itself).
QB nets are, however,
{\it complicated enough} to
be very expressive
and useful; that is, they allow
one to express numerous
quantum mechanical concepts
in a useful, practical and enlightening way.

QB nets are a very
parsimonious extension of CB nets
to quantum mechanics.
That is,
the definition of QB nets is
the smallest possible
modification of
the definition of CB nets that I can
come up with, but
enough of a modification so that
one can do proper quantum mechanics with them.
Keeping QB nets close to CB nets can
be very fruitful, because much
is already known about CB nets.
And CB nets use classical probability so
we can sharpen our classical understanding
of a problem
with them and then try to extrapolate
that understanding to QB nets.
 QB nets retain
the same structure as
CB nets and can be reduced
to them very easily,
simply by applying
the dephasing operator ``cl"
(defined below) to each node.
Thus, QB nets make the
connection to
the classical case
very direct and explicit.

\section{Basic Notation}
As usual, $\ZZ,\RR, \CC$
will denote the integers, real numbers,
and complex numbers, respectively.
For $a,b\in\ZZ$ such that $a\leq b$,
let
 $Z_{a,b}=\{a,a+1,a+2,\ldots, b\}$.

 Let
$\delta^x_{y}=\delta(x,y)$
denote the Kronecker delta function;
it equals 1 if $x=y$ and 0 if $x\neq y$.

For any matrix $M\in \CC^{p\times q}$,
$M^*$ will denote its complex conjugate,
$M^T$ its transpose, and $M^\dagger = M^{*T}$
its Hermitian conjugate.

Random variables\footnote{We will use
the term ``random variables"
in both classical and quantum physics.
Normally, random variables are
defined only in classical physics, where they are
defined to be functions from an outcome space to a
range of values.
For technical simplicity, here we
define a random variable $\rva$, in both
classical and quantum physics,
to be merely the label of a
 node in a graph, or an n-tuple
 $\rvx_K$ of such labels.
 Each node or
 random variable of a CB or QB net
 is akin to a spacetime event
 or a collection of them.} will be denoted
by underlined letters; e.g.,
$\rva$.
The (finite) set of values (states) that
$\rva$ can assume will be denoted
by $S_\rva$. Let $N_\rva=|S_\rva|$.

The probability that
$\rva=a$ will be denoted by $P(\rva=a)$
or $P_\rva(a)$, or simply by $P(a)$
if the latter will not lead to confusion
in the context it is being used.
We will use $pd(S_\rva)$ to denote
the set of all probability distributions
with domain $S_\rva$.

In quantum physics, $\rva$ has a {\it fixed,
orthonormal} basis
$\{\ket{a}_\rva:a\in S_\rva\}$
associated with it.
The vector space spanned
by this basis will be denoted by $\calh_\rva$.
Other spans of $\calh_\rva$
that are not necessarily
orthonormal will be denoted
by Greek letters with subscripts as in
$\{\ket{\psi_j}_\rva\}_{\forall j}$.
In quantum physics, instead of
probabilities $P(\rva=a)$,
we use ``probability amplitudes"
(or just ``amplitudes" for short)
$A(\rva=a)$ (also denoted by $A_\rva(a)$
or $A(a)$).
In place of $P(a)\geq 0$ and
$\sum_a P(a)=1$, one has
$\sum_a |A(a)|^2=1$.
Besides probability amplitudes,
we also use density matrices.
A density matrix
$\rho_\rva$ is a Hermitian,
non-negative, unit trace,
square matrix (or the associated
linear operator)
acting on $\calh_\rva$.
We will use $dm(\calh_\rva)$ to
denote the set of all
density matrices acting on $\calh_\rva$.

If
$\rho_{\rvx,\rva}\in dm(\calh_{\rvx,\rva})$,
and $\rho_\rvx = \tr_\rva (\rho_{\rva,\rvx})
= \sum_a \bra{a}_\rva\rho_{\rva,\rvx}\ket{a}_\rva\in dm(\calh_\rvx)$,
we will say that $\rho_\rvx$ is a
{\bf partial trace of}
$\rho_{\rvx,\rva}$.
Given a density matrix
$\rho_{\rvx_1,\rvx_2,\rvx_3,\ldots}\in
dm(\calh_{\rvx_1,\rvx_2,\rvx_3,\ldots})$,
its partial traces will be denoted
by omitting its subscripts for
the random variables that
have been traced over. For example,
$\rho_{\rvx_2} = tr_{\rvx_1,\rvx_3}
\rho_{\rvx_1,\rvx_2,\rvx_3}$.

Sometimes, when
two random variables $\rva\av{1}$ and
$\rva\av{2}$ satisfy
$S_{\rva\av{1}}=S_{\rva\av{2}}$,
we will omit the indices $\av{1}$
and $\av{2}$ and refer to both
random variables as $\rva$.
We shall do this sometimes
even if the
random variables $\rva\av{1}$
 and $\rva\av{2}$ are not identically distributed!
This notation, {\it if used with caution},
does not lead to confusion and
does avoid a lot of index clutter.

When we want to make
explicit that an operator $\Omega$
maps
states in $\calh_\rva$ to
states in $\calh_\rvb$,
we will indicate
this with a subscript (or superscript)
as $\Omega_{\rvb\leftarrow\rva}$
or as $\Omega_{\rvb|\rva}$.
In cases where
$S_\rvb=S_\rva$,
we will sometimes write
$\Omega_{\rva}$
instead of the clearer but longer
$\Omega_{\rva\leftarrow\rva}$
or $\Omega_{\rva|\rva}$.

The tensor product
symbol $\otimes$
will often be omitted.
Sometimes,
when two vectors
are being tensored,
we will list the two vectors
vertically instead of
horizontally (the latter
is more common in the literature). For example, we might write

\beq
\ket{\phi}_\rva
\otimes
\ket{\psi}_\rvb
=
\ket{\phi}_\rva
\ket{\psi}_\rvb
=
\begin{array}{c}
\ket{\phi}_\rva\\
\ket{\psi}_\rvb
\end{array}
\;.
\eeq
This doesn't lead to
confusion as long as
we indicate what vector
space each vector lives in.
(In the above example,
$\ket{\phi}_\rva$ clearly
lives in $\calh_\rva$
and $\ket{\psi}_\rvb$
in $\calh_\rvb$).

In this paper, we consider
networks (graphs) with $N$ nodes.
Each node is labeled by a random variable
$\rvx_j$,
where $j\in \zn$.
For any $J\subset \zn$,
the ordered set of random
variables $\rvx_j$ $\forall j\in J$
(ordered so that the integer indices $j$
 increase from left to right)
will be denoted by
$\rvx_{J}$.
For example, $
\rvx_{\{2,4\}}=(\rvx_2,\rvx_4)$.
We will often call the values that
$\rvx_{J}$
can assume $x_{J}$.
For example, $
x_{\{2,4\}}=(x_2,x_4)$.
We will often abbreviate
$\rvx_{\zn}$
by just
$\rvxdot$\;\;.
We will often call the values that
$\rvxdot$
can assume $x.$\;\;.

\section{The Sandbox and its Dual}

For any expressions $\Omega(x)$ and $\rho$
for which this makes sense,
we will use the shorthand notation:

\beq
\sandb{\Omega(x)}
\rho
\sandb{
\begin{array}{c}
\hc \\
x\rarrow x'
\end{array}
}
=
\sandb{\Omega(x)}
\rho
\sandb{
\Omega^\dagger(x')
}
\;.
\eeq
Here ``h.c." is an abbreviation
of ``hermitian conjugate".
We will usually use this notation with
$\rho=1$. This notation
is especially useful when $\Omega(x)$
is a long expression
and we want to avoid writing it twice.
We will
refer to the space inside the set
of square brackets to the left (resp., right)
of $\rho$ as {\bf the sandbox}
(resp., {\bf its dual or mirror sandbox}).

\section{The Meta State}
QB nets for pure quantum states
were first defined in Ref.\cite{Tuc97}.
A QB net for
a pure state consists of a
directed acyclic graph (DAG)
and a transition matrix (a complex matrix)
assigned to each node of the graph.
The transition matrices must satisfy
certain requirements.
An example of such a
pure state QB net is:

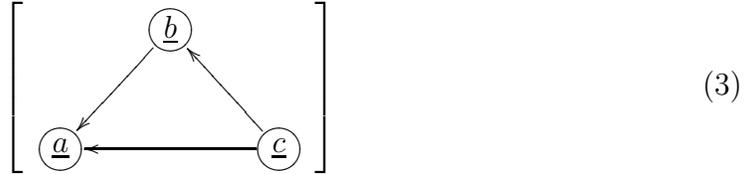
\beq
\sandb{
\entrymodifiers={++[o][F-]}
\xymatrix{
*{}&\rvb\ar[dl]&*{} \\
\rva&*{}&\rvc\ar[ul]\ar[ll]
}}
\eeq
If $a\in S_\rva, b\in S_\rvb, c\in S_\rvc$,
$A_{\rva|\rvb,\rvc}(a|b,c)$
 is the transition
matrix associated with node $\rva$,
$A_{\rvb|\rvc}(b|c)$ is the transition
matrix for
node $\rvb$, and $A_\rvc(c)$
is the transition
matrix for node $\rvc$.
We must have
\begin{subequations}\label{eq-row-sums}
\beq
\sum_a |A_{\rva|\rvb,\rvc}(a|b,c)|^2=1
\;,
\eeq
\beq
\sum_b |A_{\rvb|\rvc}(b|c)|^2=1
\;,
\eeq
\beq
\sum_c |A_{\rvc}(c)|^2 = 1
\;.
\eeq
\end{subequations}
Define the total probability
amplitude $A_{\rva,\rvb,\rvc}(a,b,c)$
by

\beq
A_{\rva,\rvb,\rvc}(a,b,c) =
A_{\rva|\rvb,\rvc}(a|b,c)
A_{\rvb|\rvc}(b|c)
A_{\rvc}(c)
\;.\label{eq-def-tot-a}
\eeq
Note that
Eqs.(\ref{eq-row-sums}) imply

\beq
\sum_{a,b,c} |A_{\rva,\rvb,\rvc}(a,b,c)|^2 = 1
\;.\label{eq-bnet-unitarity}
\eeq

Henceforth, we will sometimes
omit the node subscripts from
the probability amplitudes.
For example, we might
use $A(a|b,c)$
instead of
$A_{\rva|\rvb,\rvc}(a|b,c)$,
if no confusion will arise.
This is analogous
to probability theory, where
we often
use $P(a|b,c)$
instead of
$P_{\rva|\rvb,\rvc}(a|b,c)$
or $P(\rva=a|\rvb=b,\rvc=c)$
for a probability.

More generally, suppose the graph
has $N$ nodes $\rvx_1,\rvx_2, \ldots,\rvx_N$.
For $j\in \zn$,
a node
$\rvx_j$ with possible states
$x_j\in \sts{\rvx_j}$
and with parent nodes
$\rvx_{pa(\rvx_j)}$ where
$pa(\rvx_j)\subset \zn$,
has a transition matrix
$A(x_j|x_{pa(\rvx_j)})$ which satisfies

\beq
\sum_{x_j\in \sts{\rvx_j}}
|A(x_j|x_{pa(\rvx_j)})|^2=1
\;\label{eq-aj-norms}
\eeq
for all $x_{pa(\rvx_j)}\in \sts{\rvx_{pa(\rvx_j)}}$.
Let $x_.=(x_1,x_2, \ldots,x_N)$
If the total amplitude $A(x.)$
is defined by

\beq
A(x.) = \prod_{j\in \zn}
A(x_j|x_{pa(\rvx_j)})
\;,
\eeq
then Eqs.(\ref{eq-aj-norms}) imply

\beq
\sum_{x.}|A(x.)|^2=1
\;.
\eeq

Given any transition matrix
of the form $A(r|\vec{c})$,
call $r$ the row index and $\vec{c}$
the column indices. Call all $A(r|\vec{c})$
entries
with any $r$ but
fixed $\vec{c}$,
a column vector
of the transition matrix.
Eqs.(\ref{eq-row-sums}) say that
each column vector of a transition
matrix is normalized. The column vectors
may also be mutually orthogonal,
in
which case
we say that the column vectors are orthonormal. For
example, the transition
matrix for node $\rva$
in the QB net above
might also satisfy:

\beq
\sum_a \sandb{A_{\rva|\rvb,\rvc}(a|b,c)}
\sandb{\begin{array}{c}
\hc\\ b,c\rarrow b',c'
\end{array}}
= \delta_{b}^{b'}\delta_{c}^{c'}
\;.
\eeq
The isometry nodes
defined below are
another example of a case
where the column vectors of the transition matrix
are orthonormal. In general, it
is not necessary that the column vectors
 be orthonormal.
For example, for the marginalizer
nodes defined below, they aren't.
What is always necessary is
that the total amplitude be normalized,
as in Eq.(\ref{eq-bnet-unitarity}), so as to
enforce the ``unitarity" of quantum
mechanics.

The meta state of a QB net was
first defined in Ref.\cite{Tuc07}.
A {\bf meta ket state}
is a pure quantum state
represented as a ket or as a QB net.
For example:

\beqa
\ket{\psi_{\rm meta}}_{\rva,\rvb,\rvc}&=&
\sandb{
\entrymodifiers={++[o][F-]}
\xymatrix{
*{}&\rvb\ar[dl]&*{} \\
\rva&*{}&\rvc\ar[ul]\ar[ll]
}}
\\
&=& \sum_{a,b,c} A_{\rva,\rvb,\rvc}(a,b,c)
\begin{array}{c}
\ket{a}_\rva\\
\ket{b}_\rvb\\
\ket{c}_\rvc
\end{array}
\;,
\eeqa
where $A(a,b,c)$
is defined by Eq.(\ref{eq-def-tot-a}).
We assume
the states $\{\ket{a}_\rva\}_{\forall a}$
are orthonormal, and likewise
for  $\{\ket{b}_\rvb\}_{\forall b}$
and $\{\ket{c}_\rvc\}_{\forall c}$.
Note that each
node of the QB net has its own ket
and its own index that is summed over
(i.e., bound). For example, node $\rvb$
has ket $\ket{b}_\rvb$ and bound index $b$.

The projection operator
of a meta ket defines
a density matrix which we will call the
{\bf meta density matrix} of the protocol
under consideration. For example,
the meta density matrix of the
above meta ket is given by:

\beqa
(\rho_{\rm meta})_{\rva,\rvb,\rvc} &=&
\sandb{
\ket{\psi_{\rm meta}}_{\rva,\rvb,\rvc}
}\sandb{\hc}\\
&=&
\sandb{
\entrymodifiers={++[o][F-]}
\xymatrix{
*{}&\rvb\ar[dl]&*{} \\
\rva&*{}&\rvc\ar[ul]\ar[ll]
}
}
\sandb{\hc}
\;.
\eeqa

\section{Generic Nodes}
We find it convenient to define
certain special, generic types
of nodes.

{\bf Marginalizer nodes} were first
defined in Ref.\cite{Tuc97}.
In the current version of
Quantum Fog, marginalizer
nodes are usually denoted by black bullets,
whereas non-marginalizer nodes
are denoted by larger colored circles.
In this paper, we will represent
marginalizer nodes by writing
a small delta near them. This
node ``decoration" or subscript is easy to
draw by hand and also easy
for the eye to spot.
For example:
\beq
\ket{\psi}=
\sandb{
\entrymodifiers={+++++[o][F-]}
\xymatrix{
\rva_{\av{2}}&*{}&*{}\\
*{}&*{}&(\rva_{\av{1}},\rvb_{\av{1}})
\ar[ull]_>{\delta}
\ar[dll]^>{\delta} \\
\rvb_{\av{2}}&*{}&*{}
}
}
\;,\label{eq-cluttered}
\eeq
where $\rva_{\av{1}}$ and $\rva_{\av{2}}$
have the same state space, call it $S_\rva$.
 Likewise,
$\rvb_{\av{1}}$ and $\rvb_{\av{2}}$
have the same state space, call it $S_\rvb$.
Note that the subscripts $\av{1}$ and $\av{2}$
are acting like a ``time" index along a sort
of timeline.
For all $a,a'\in S_\rva$ and $b,b'\in S_\rvb$,

\beq
\begin{array}{l}
A_{\rva_{\av{2}}|\rva_{\av{1}},\rvb_{\av{1}}}(a|a',b')=
\delta_a^{a'}\\
A_{\rvb_{\av{2}}|\rva_{\av{1}},\rvb_{\av{1}}}(b|a',b')=
\delta_b^{b'}
\end{array}
\;.
\eeq
Thus some column vectors of
a marginalizer node are
equal to each other.
To avoid index clutter,
we will sometimes omit the indices
$\av{1}$ and $\av{2}$
from the graph of Eq.(\ref{eq-cluttered}),
and draw instead
the following graph:

\beq
\ket{\psi}=
\sandb{
\entrymodifiers={++[o][F-]}
\xymatrix{
\rva&*{}&*{}\\
*{}&*{}&(\rva,\rvb)
\ar[ull]_>{\delta}
\ar[dll]^>{\delta} \\
\rvb&*{}&*{}
}
}
\;
\eeq

{\bf Grounded nodes} are root nodes (i.e., nodes
with no incoming arrows,
only outgoing ones)
which
have a deterministic
probability amplitude (i.e., an amplitude
that equals 1 for just one of the
possible states of the node and zero for
all the other states).
Grounded nodes will be indicated
by writing a zero
near them.
Here is
an example of a QB net
with a grounded node:

\beq
\sandb{
\entrymodifiers={++[o][F-]}
\xymatrix{
*{}&*{}&\rva\ar[dll]\\
(\rva,\rvb)&*{}&*{}\\
*{}&*{}&\rvb\ar[ull]_<{0}
}
}
\;,
\eeq
where\footnote{
We are assuming that $0\in S_\rvb$.
It doesn't matter if the
amplitude $A(b)$
of node $\rvb$ equals $\delta_b^0$,
or $\delta_b^{b_0}$ where $b_0\in S_\rvb$.
Either way, it's still a grounded node.
An alternative to writing
a zero next to a node
to indicate that it's grounded
might be
writing instead
the letters ``grd"
or the electrical symbol for a ground.}

\beq
A(b)= \delta_b^0
\;,
\eeq
for all $b\in S_\rvb$.

\section{Isometries}

Consider the following
QB net

\beq
\sandb{
\entrymodifiers={++[o][F-]}
\xymatrix{
\rvb&\rva\ar[l]
}
}
\;,
\eeq
where $A(b|a)$ satisfies

\beq
\sum_{b\in S_\rvb}\sandb{A(b|a)}
\sandb{
\begin{array}{c}\hc \\
a\rarrow a'
\end{array}
}
=
\delta_{a}^{a'}
\;,\label{eq-simple-or-no-iso}
\eeq
for all $a,a'\in S_\rva$.
The node $\rvb$ in the above
QB net is called
an {\bf isometry node}, or just an isometry for
simplicity.

Eq.(\ref{eq-simple-or-no-iso})
is saying that the column vectors of
the transition matrix $A(b|a)$ are orthonormal.
This is only possible if $N_\rvb\geq N_\rva$.
If $N_\rvb= N_\rva$ (i.e., transition
matrix is square), then the transition
matrix is unitary. If $N_\rvb> N_\rva$
(i.e., transition
matrix is rectangular with more rows than columns),
then we can use the well-known, so called
Gram-Schmidt procedure to add more columns
to the transition matrix (``extend it")
to produce a unitary matrix.

Since $N_\rvb\geq N_\rvb$
and the sets $S_\rvb$, $S_\rva$ are finite,
we may assume without loss
of generality that $S_\rvb \supset S_\rva$.
For every $b,A\in S_\rvb$, let\footnote{We are
using the symbol $A$
both for an element $A$ of $S_\rvb$,
and for amplitudes $A(\cdot)$.
It's easy to tell which usage
is intended in each instance,
so this should cause no confusion.}

\beq
A(b|A) =\left\{
\begin{array}{l}A(b|a)\;\;
\mbox{if}\;\;A=a\in S_\rva\\
\mbox{given by Gram-Schmidt procedure}\;\;{\rm if}\;\;A\in S_\rvb-S_\rva
\end{array}\right.
\;.
\eeq
Then

\beq
\sum_{b\in S_\rvb} \sandb{A(b|A)}
\sandb{
\begin{array}{c}\hc\\A\rarrow A'\end{array}
}
=\delta^{A}_{A'}
\;
\eeq
for all $A,A'\in S_\rvb$. Thus,

\beq
A(b|A)=
\bra{b}U_{\rvb}\ket{A}
\;,
\eeq
where $U_{\rvb}$ is unitary.

Here is a
 pictorial
representation,
in terms of QB nets, of
the procedure
just outlined for
extending an isometry to
a unitary matrix:

\beq
\sandb{
\entrymodifiers={++[o][F-]}
\xymatrix{
\rvb&\rva\ar[l]
}
}
\rarrow
\sandb{
\entrymodifiers={++[o][F-]}
\xymatrix{
\rvb&\rv{A}\ar[l]
}
}
\;,
\eeq
where $S_\rvb=S_\rv{A}\supset S_\rva$.

Consider the following
QB net

\beq
\sandb{
\entrymodifiers={++[o][F-]}
\xymatrix{
(\rva,\rvb)&\rva\ar[l]
}
}
\;,
\eeq
where $A(a,b|a')$ satisfies

\beq
\sum_{a\in S_\rva}
\sum_{b\in S_\rvb}\sandb{A(a,b|a')}
\sandb{
\begin{array}{c}\hc \\
a'\rarrow a''
\end{array}
}
=
\delta_{a'}^{a''}
\;,
\eeq
for all $a',a''\in S_\rva$.
The node $(\rva, \rvb)$ in the above
QB net is a special case
of the general isometry node presented previously.
Just as in the case of a general isometry,
the transition matrix $A(a,b |a')$
can be extended to a
unitary matrix. Assume $0\in S_\rvb$.
If we define

\beq
A(a, b|a', b'=0)\;=\; A(a,b|a')
\;,
\eeq
then we can find a unitary matrix $U_{\rva,\rvb}$ such that,
for all $a,a'\in S_\rva$ and
$b,b'\in S_\rvb$,

\beq
A(a, b|a', b')=
\begin{array}{ccc}
\bra{a}_\rva&&\ket{a'}_\rva\\
& U_{\rva,\rvb} & \\
\bra{b}_\rvb&&\ket{b'}_\rvb
\end{array}
\;.
\eeq

Here is a
 pictorial
representation,
in terms of QB nets, of
the procedure
just outlined for
extending an isometry to
a unitary matrix:

\beq
\sandb{
\entrymodifiers={++[o][F-]}
\xymatrix{
(\rva,\rvb)&\rva\ar[l]
}
}
=
\sandb{
\entrymodifiers={++[o][F-]}
\xymatrix{
*{}&*{}&\rva\ar[dll]\\
(\rva,\rvb)&*{}&*{}\\
*{}&*{}&\rvb\ar[ull]_<{0}
}
}
\rarrow
\sandb{
\entrymodifiers={++[o][F-]}
\xymatrix{
*{}&*{}&\rva\ar[dll]\\
(\rva,\rvb)&*{}&*{}\\
*{}&*{}&\rvb\ar[ull]
}
}
\;
\eeq

\section{Freeing a Bound Index}
We've defined the
QB net corresponding to
the meta ket state
as having for each node $\rvb$: (1)
an index $b$ that is
summed over (bound), and (2)
a ket $\ket{b}_\rvb$.
One can free an index $b$
of a node $\rvb$
of a meta ket state
by multiplying that
node by $\bra{b}$
or $\ket{b}\bra{b}$.
One can use QB nets
 to represent these
two operations. For example,
if the meta state is,

\beq
\sandb{
\entrymodifiers={++[o][F-]}
\xymatrix{
\rvb&\rva\ar[l]
}
}
=
\sum_{a,b} A(b|a)A(a)\ket{b}_\rvb\ket{a}_\rva
\;,
\eeq
then

\beq
\sandb{
\entrymodifiers={++[o][F-]}
\xymatrix{
*{}&
\rvb\ar@{}[l]_<<{\bra{b}}&\rva\ar[l]
}
}
=
\sum_{a} A(b|a)A(a)\ket{a}_\rva
\;,
\eeq
and

\beq
\sandb{
\entrymodifiers={++[o][F-]}
\xymatrix{
*{}&
\rvb\ar@{}[l]_<<<<{\ket{b}\bra{b}}&\rva\ar[l]
}
}
=
\ket{b}_\rvb\sum_{a} A(b|a)A(a)\ket{a}_\rva
\;.
\eeq

\section{Classical Communication}
Quantum information theorists
call ``classical communication"
the act of measuring an observable
at one event and then using
the result of that measurement
to start a new event.
Classical communication
can be represented using
QB nets. For example,
if $S_\rvc= S_\rvb$, then

\beq
\sandb{
\entrymodifiers={++[o][F-]}
\xymatrix{
\rvd&\rvc\ar[l]_<<{\bra{b}}&\rvb\ar@{}[l]_<<{\bra{b}}&\rva\ar[l]
}
}
=
\sum_{d,a}
A_{\rvd|\rvc}(d|b)A_\rvc(b)
A_{\rvb|\rva}(b|a)A_\rva(a)
\ket{d}_\rvd\ket{a}_\rva
\;.
\eeq

\section{(Coherent or Incoherent)-(Scalar or Vector) Sums}
For any
$\rho_{\rvb,\rva}\in dm(\calh_{\rvb,\rva})$,
define

\beq
\tr_\rvb(\rho_{\rvb,\rva}) = \rho_\rva =
\sum_b \sandb{\bra{b}_\rvb}
\;\;\rho_{\rvb,\rva}\;\;
\sandb{\hc}
\;,
\eeq

\beq
{\rm cl}_\rvb(\rho_{\rvb,\rva}) =
\rho_{\rvb_{cl},\rva} =
\sum_b \sandb{\ket{b}_\rvb\bra{b}_\rvb}
 \;\;\rho_{\rvb,\rva}\;\;
\sandb{\hc}
\;,
\eeq

\beq
{\rm sl}_\rvb(\rho_{\rvb,\rva}) =
\rho_{\cancel{\rvb}\rva} =
 \;\;\sandb{\sum_b\bra{b}_\rvb}\;\;
\;\;\rho_{\rvb,\rva}\;\;
\sandb{\hc}
\;.
\eeq
Note that

\beq
\tr_\rvb(1) = N_\rvb,\;\;
{\rm cl}_\rvb(1) = 1,\;\;
{\rm sl}_\rvb(1) = N_\rvb
\;.
\eeq
Note also that the product
of any two operators
in the set $F=\{1, \tr_\rvb, {\rm cl}_\rvb,
 {\rm sl}_\rvb\}$
can be expressed in terms of a single
one of them. For example,

\beq
\tr_\rvb {\rm cl}_\rvb(\rho_{\rvb,\rva})=
\tr_\rvb(\rho_{\rvb,\rva})
\;,\;\; \tr_\rvb \tr_\rvb(\rho_{\rvb,\rva})=
N_\rvb \tr_\rvb(\rho_{\rvb,\rva})
\;,
\eeq
etc.. Hence, a product of any number of
the operators in $F$ can be expressed in terms
of a single one of them.\footnote{More formally,
if we define
$c= {\rm cl}_\rvb,$
$\sigma = {\rm sl}_\rvb/N_\rvb,$
$\tau = \tr_\rvb/N_\rvb$, and
$F= \{1, c,\sigma,\tau\}$,
then it is easy to check that for all $f\in F$,
$f\tau = \tau$, $f\sigma=\sigma$,
and $f c=\left\{\begin{array}{l}
c\;\;{\rm if}\;\;f\in\{1,c\}\\
\tau\;\;{\rm otherwise}
\end{array}\right.$.
Although $F$ is
closed under composition, it
is not a group.
This is not surprising since
$c,\rho,\sigma$ are irreversible
transformations.}

For each node  $\rvb$ of
a meta density matrix,
there is an index $b$
that is summed over and a
ket $\ket{b}_\rvb$.
Furthermore, the $\sum_b$ is inside
the sandbox, so we say
that it's a {\bf coherent sum}.
Because the term being
summed (i.e., the summand) includes the ket
$\ket{b}_\rvb$, we say
it's a {\bf vector sum}.

If
the $\sum_b$ were outside the
sandbox (and index $b$
appeared in both the sandbox and its
dual), we would call it an
{\bf incoherent sum}.
If the summand did not
include $\ket{b}_\rvb$,
we would call it a {\bf scalar sum}.
The operators $\tr_\rvb(\cdot)$,
${\rm cl}_\rvb(\cdot)$, ${\rm sl}_\rvb(\cdot)$
act on the meta density matrix
to change the
coherent-vector sum over a node $\rvb$
to an incoherent-scalar, or
an incoherent-vector,
 or
a coherent-scalar sum. Let's
illustrate this with an example.

\begin{itemize}
\item Consider the following
meta density matrix as an example. Note that
in this meta density matrix, for the
random variable $\rvb$, there is a
 {\bf coherent-vector sum} over the index $b$.
\beqa
\rho_{\rvb, \rva}
&=&
\sandb{\sum_{a,b} A(b|a)A(a)\ket{b}_\rvb\ket{a}_\rva}
\sandb{\hc}\\
&=&
\sandb{
\entrymodifiers={++[o][F-]}
\xymatrix{
\rvb&\rva\ar[l]
}
}
\sandb{\hc}
\;
\eeqa

\item ``Tracing" (i.e., taking a partial trace of)
the random variable $\rvb$
means doing an {\bf incoherent-scalar sum}
over the index $b$.

\beqa
\tr_\rvb(\rho_{\rvb,\rva}) =\rho_{\rva}
&=&
\sum_b\sandb{\sum_{a} A(b|a)A(a)\ket{a}_\rva}
\sandb{\hc}\\
&=&
\sandb{
\entrymodifiers={++[o][F-]}
\xymatrix{
\rvb&\rva\ar[l]_>>{tr}
}
}
\sandb{\hc}
\\
&=&\sum_b
\sandb{
\entrymodifiers={++[o][F-]}
\xymatrix{
*{}& \rvb\ar@{}[l]_<<{\bra{b}}&\rva\ar[l]
}
}
\sandb{\hc}
\;
\eeqa

\item ``Classicizing", or ``Making classical" the
random variable $\rvb$
means doing an {\bf incoherent-vector sum} over
the index $b$. (This operation
is also sometimes
described as ``dephasing"
 because we are throwing away
some off-diagonal
terms).

\beqa
{\rm cl}_\rvb(\rho_{\rvb, \rva})=
\rho_{\rvb_{cl}, \rva}
&=&
\sum_b\sandb{\sum_{a} A(b|a)A(a)\ket{b}_\rvb\ket{a}_\rva}
\sandb{\hc}
\\ &=&
\sandb{
\entrymodifiers={++[o][F-]}
\xymatrix{
\rvb&\rva\ar[l]_>>{cl}
}
}
\sandb{\hc}
\\ &=&\sum_b
\sandb{
\entrymodifiers={++[o][F-]}
\xymatrix{
*{}&\rvb\ar@{}[l]_<<<<{\ket{b}\bra{b}}&\rva\ar[l]
}
}
\sandb{\hc}
\;
\eeqa

\item ``Slashing"
the random variable $\rvb$
means doing a {\bf coherent-scalar sum} over
the index $b$.
\beqa
{\rm sl}_\rvb(\rho_{\rvb, \rva})=
\rho_{\cancel{\rvb}, \rva}
&=&
\sandb{\sum_{a,b} A(b|a)A(a)\ket{a}_\rva}
\sandb{\hc}
\\ &=&
\sandb{
\entrymodifiers={++[o][F-]}
\xymatrix{
\cancel{\rvb}&\rva\ar[l]
}
}
\sandb{\hc}
\\ &=&
\sandb{\sum_b
\entrymodifiers={++[o][F-]}
\xymatrix{
*{}&\rvb\ar@{}[l]_<<{\bra{b}}&\rva\ar[l]}
}
\sandb{\hc}
\;
\eeqa
\end{itemize}

Note that the operators in $F$ are all
irreversible transformations
(except for the 1).
The meta state is
truly ``at the top
of the food chain":
Once the operators
in $F$ take the meta state
to something else, no
operator or combination of operators
in $F$ can bring back the
same meta state.

\section{Ensembles, Purification}
An {\bf ensemble} is a set $\{\sqrt{w_j}
\ket{\psi_j}_\rvx\}_{\forall j}$
where the weights $w_j$ are non-negative numbers
that sum to 1, and for all $j$,
the states $\ket{\psi_j}_\rvx\in \calh_\rvx$
are normalized but they are not necessarily
mutually orthogonal.  The density matrix
for this ensemble is

\beq
\rho_\rvx = \sum_j w_j
\sandb{\ket{\psi_j}_\rvx}\sandb{\hc}
\;.\label{eq-rhox}
\eeq
Define two ensembles as being
equivalent if they have the same density matrix.
This defines an equivalence relation.
Elements of the same equivalence class
are physically indistinguishable.

The density matrix
Eq.(\ref{eq-rhox}) can be {\bf purified},
meaning that it can
be expressed as a partial trace
of a pure state. One way
of doing this is as follows.
Clearly, $\rho_\rvx$ also equals

\beq
\rho_\rvx =
\tr_\rvj
\sandb{
\begin{array}{r}
\sum_x \ket{x}\av{x|\psi_j}_\rvx\\
\sum_j \sqrt{w_j}\ket{j}_\rvj
\end{array}
}
\sandb{\hc}
\;.
\eeq
Thus

\beq
\rho_\rvx =\tr_\rvj
\sandb{\ket{\psi}_{\rvx,\rvj}}
\sandb{\hc}
\;,
\eeq
where

\beq
\ket{\psi}_{\rvx,\rvj} =
\sandb{\sum_{x,j}A(x,j)
\begin{array}{c}
\ket{x}_\rvx\\
\ket{j}_\rvj
\end{array}
}
=
\sandb{
\entrymodifiers={++[o][F-]}
\xymatrix{\rvx&\rvj\ar[l]}
}
\;,
\eeq
where

\beq
A(x,j) = A(x|j)A(j)
\;,
\eeq
and

\beq
A(x|j) = \av{x|\psi_j}
,\;\;\;
A(j) = \sqrt{w_j}
\;.
\eeq
\section{Measurement Superoperators}
\label{sec-measurements}

A {\bf superoperator}
is
a linear operator that maps $dm(\calh_\rva)$
into $dm(\calh_\rvb)$.

A {\bf measurement}
is defined as a set
$\{K_\mu|\mu\in S_{\rv{\mu}}\}$
of operators $K_\mu$ called {\bf Krauss
operators} that map states
in $\calh_\rva$ to states in $\calh_\rvb$.
We assume $N_\rva\leq N_\rvb$.
($N_\rva = |S_\rva| = dim(\calh_\rva)$ and the
same for $\rvb$).
The Krauss operators must also satisfy:

\beq
\sum_\mu K^\dagger_\mu K_\mu = 1
\;.
\eeq

Each Krauss operator $K_\mu$
can be used to define
a superoperator $\$_\mu(\cdot)$
as follows. Let
$\rho_\rva\in dm(\calh_\rva)$,
and, for each $\mu$, let
$\sigma_{\rvb|\mu}\in dm(\calh_\rvb)$. Then define
the {\bf measurement superoperator}
$\$_\mu(\cdot)$ by

\beq
\$_\mu(\rho_\rva)=
\rho_{\rvb|\mu} =
\frac{K_\mu \rho_\rva K^\dagger_\mu}
{P(\mu)}
\;,
\eeq
where

\beq
P(\mu) = \tr_\rvb( K_\mu \rho_\rva K^\dagger_\mu)=
\tr_\rva(K^\dagger_\mu K_\mu \rho_\rva )
\;.
\eeq
Note that the $P(\mu)$ are non-negative and

\beq
\sum_\mu P(\mu)=1
\;.
\eeq
Note also that

\beq
\tr_\rvb(\rho_{\rvb|\mu})=1
\;
\eeq
for all $\mu$.

A {\bf von Neumann measurement} (for instance,
$K_\mu = \ket{\mu}\bra{\mu}$)
is a measurement
$\{K_\mu\}_{\forall \mu}$ that satisfies:

\beq
K^\dagger_\mu = K_\mu,\;\;
K_\mu K_{\mu'} = \delta_\mu^{\mu'},\;\;
\sum_\mu K_\mu = 1
\;.
\eeq

Here are some other examples of measurements
(you can check that $\sum_a K_a^\dagger K_a =1$
for each example)
\begin{itemize}
\item Tracing: $K_a = \bra{a}_\rva$
\item Making a node classical (i.e., dephasing  it):
$K_a = \ket{a}_\rva\bra{a}_\rva$
\item Classical (incoherent) communication:
$K_a = \ket{a}_{\rvb}\bra{a}_{\rva}$, where $S_\rva=S_\rvb$.
\item Coherent communication
(only one Krauss operator):
$K = \sum_a
\ket{a}_{\rvb}\bra{a}_{\rva}$, where $S_\rva=S_\rvb$.
\end{itemize}

A measurement $\{K_\mu\}_{\forall \mu}$ can
be extended to a unitary operator
as follows. For every $b\in S_\rvb$,
$\mu\in S_{\rv{\mu}}$,
$a\in S_\rva$, define

\beq
A(b,\mu|a)=\av{b|K_\mu|a}
\;.
\eeq
Since for all $a,a'\in S_\rva$,

\beq
\sum_{b,\mu} \sandb{A(b,\mu|a)}
\sandb{
\begin{array}{c}\hc\\a\rarrow a'\end{array}
}
=
\sum_{b,\mu} \av{b|K_\mu|a}
\av{a'|K^\dagger_\mu|b}
=\delta^{a}_{a'}
\;,
\eeq
$A(b,\mu|a)$ defines an isometry.
Assume $0\in S_{\rv{\mu}}$. Let

\beq
A(b,\mu|a,\mu'=0)=A(b,\mu|a)
\;.
\eeq
Since $N_\rva\leq N_\rvb$,
we can use Gram Schmidt
to find a
 unitary operator $U_{\rvb,\rv{\mu}}$ such that

\beq
A(b,\mu|A, \mu')=
\begin{array}{lcr}
\bra{b}_\rvb & & \ket{A}_{\rv{A}}\\
& U_{\rvb,\rv{\mu}} & \\
\bra{\mu}_{\rv{\mu}} & & \ket{\mu'}_{\rv{\mu}}
\end{array}
\;
\eeq
for all $b,A\in S_\rvb=S_{\rv{A}}$ and
$\mu,\mu'\in S_{\rv{\mu}}$.

\section{RINNO (POVM)}
A  POVM, which
I prefer to call a RINNO, is
a Resolution of the
Identity by Non Negative Operators. Thus,
a RINNO
$\{R_\mu\}_{\forall \mu}$
 satisfies

\beq
\sum_\mu R_\mu = 1,\;\; R_\mu \geq 0
\;.\label{eq-rinno-def}
\eeq
(Each $R_\mu$ is a
square matrix. A square matrix
$M$ is said to be non-negative,
or said to satisfy $M\geq 0$,
if $v^\dagger M v\geq 0$
for all complex column vectors $v$.)

Suppose $\rho_\rva \in dm(\calh_\rva)$,
and for each $\mu$,
 $R_\mu$ maps $\calh_\rva$ into
itself. For each $\mu$, define

\beq
P(\mu) = \tr_\rva(R_\mu \rho_\rva)
\;.
\eeq
By
Eqs.(\ref{eq-rinno-def}), the $P(\mu)$
are non-negative and satisfy $\sum_\mu P(\mu)=1$.

A RINNO
$\{R_\mu\}_{\forall \mu}$
can be constructed
from a measurement $\{K_\mu\}_{\forall \mu}$
by setting

\beq
R_\mu = K^\dagger_\mu K_\mu
\;\label{eq-rinno-k}
\eeq
for each $\mu$.
The definition of a measurement
$\{K_\mu\}_{\forall \mu}$
and Eqs.(\ref{eq-rinno-k})
 imply Eqs.(\ref{eq-rinno-def}).

\section{Channel Superoperators}

Suppose $\{K_\mu|\mu\in S_{\rv{\mu}}\}$
is a measurement with
Krauss operators
$K_\mu:\calh_\rva\rarrow\calh_\rvb$.
Let
$\rho_\rva\in dm(\calh_\rva)$
and
$\sigma_{\rvb}\in dm(\calh_\rvb)$. Then define the
{\bf channel superoperator} $\$(\cdot)$ by\footnote{
Krauss showed that for any
superoperator $\$(\cdot)$,\;\;
$\$(\cdot)$ is a channel
superoperator
iff $\$(\cdot)$ is ``completely positive".
}

\beq
\$(\rho_\rva)=
\sigma_\rvb=
\sum_\mu K_\mu \rho_\rva K_\mu^\dagger
\;.\label{eq-def-channel}
\eeq
Note that a channel superoperator
is a weighted sum
of measurement superoperators (i.e.,
$\$ = \sum_\mu P(\mu) \$_\mu$).

$\rho_\rva$ can always be expressed as

\beq
\rho_\rva = \sum_j w_j
\sandb{\ket{\psi_j}_\rva}\sandb{\hc}
\;,\label{eq-rhoa}
\eeq
where the weights $\{w_j\}_{\forall j}$
are non-negative numbers that sum to one,
and the states
$\{\ket{\psi_j}_\rva\}_{\forall j}$ are
all normalized but
not necessarily mutually orthogonal.
Note that for all $b,b'\in S_\rvb$,

\beqa
\av{b|\sigma_\rvb|b'} &=& \sum_{\mu,j}
\sandb{\av{b|K_\mu|\psi_j}\sqrt{w_j}}
\sandb{\begin{array}{c}\hc\\
b\rarrow b'\end{array}}
\\
&=&\sum_{\mu,j}
\sandb{
\begin{array}{ccc}
\bra{b}_\rvb&&\ket{\psi_j}_{\rv{A}}\\
& U_{\rvb,\rv{\mu}} & \\
\bra{\mu}_\rv{\mu}&&\ket{0}_\rv{\mu}
\end{array}\sqrt{w_j}
}\sandb{\begin{array}{c}
\hc\\b\rarrow b'\end{array}}
\;\label{eq-sigb}
\eeqa
where, as discussed in
Section \ref{sec-measurements},
$U_{\rvb,\rv{\mu}}$
is a unitary matrix that
extends the measurement
$\{K_\mu|\mu\in S_{\rv{\mu}}\}$.

Eq.(\ref{eq-rhoa})
for $\rho_\rva$ and Eq.(\ref{eq-sigb})
for $\sigma_\rvb$
can be represented
as follows in terms of
QB nets:

\beq
\rho_\rva= \tr_\rvj
\sandb{
\entrymodifiers={++[o][F-]}
\xymatrix{\rva&\rvj\ar[l]}
}
\sandb{\hc}
\;,\label{eq-net-rhoa}
\eeq

\beq
\sigma_\rvb= \tr_{\rv{\mu},\rvj}
\sandb{
\entrymodifiers={+++[o][F-]}
\xymatrix{
\cancel{\rvb}&*{}&\cancel{\rv{A}}\ar[dl]&\rvj\ar[l]\\
*{}&(\rvb,\rv{\mu})
\ar[ul]_>>{\delta}
\ar[dl]^>>{\delta}&*{}&*{}\\
\cancel{\rv{\mu}}&*{}&\cancel{\rv{\mu}}
\ar[ul]^<{0}&*{}
}
}
\sandb{\hc}
\;,\label{eq-net-sigb}
\eeq
where

\beq
A(\mu) = \delta_\mu^0
\;,
\eeq

\beq
A(j) = \sqrt{w_j}
\;,
\eeq

\beq
A(A|j) = \left\{\begin{array}{l}
\av{A|\psi_j}\;\;{\rm if}\;\; A\in S_\rva\\
0\;\;{\rm if}\;\; A\in S_\rvb-S_\rva
\end{array}\right.
\;,
\eeq

\beq
A(b,\mu|A,\mu')=
\begin{array}{ccc}
\bra{b}_\rvb&&\ket{A}_{\rv{A}}\\
& U_{\rvb,\rv{\mu}} & \\
\bra{\mu}_\rv{\mu}&&\ket{\mu'}_\rv{\mu}
\end{array}
\;,
\eeq

\beq
A(b|b',\mu')=\delta_b^{b'}
\;,
\eeq

\beq
A(\mu|b',\mu')=\delta_\mu^{\mu'}
\;.
\eeq
\section{Complementary Channel}

The channel superoperator $\$(\cdot)$
given by Eq.(\ref{eq-def-channel})
can be used to define
a {\bf complementary channel} superoperator
$\$'(\cdot)$. If $\$(\cdot)$
is generated using a measurement
$\{K_\mu\}_{\forall \mu}$, then
we can find a unitary operator
$U_{\rvb,\rv{\mu}}$ such that

\beq
K_\mu^{\rvb\leftarrow\rv{A}} =
\begin{array}{lcr}
\frac{\;\;\;}{}& & \frac{\;\;\;}{}\\
& U_{\rvb,\rv{\mu}} &\\
\bra{\mu}_{\rv{\mu}} & & \ket{0}_{\rv{\mu}}
\end{array}
\;.
\eeq
Now define a measurement
$\{L_b\}_{\forall b}$
using the same unitary operator
$U_{\rvb,\rv{\mu}}$:

\beq
L_b^{\rv{\mu}\leftarrow\rv{\mu}} =
\begin{array}{lcr}
\bra{b}_\rvb & & \ket{0}_{\rv{A}}\\
& U_{\rvb,\rv{\mu}} & \\
\frac{\;\;\;}{}& & \frac{\;\;\;}{}
\end{array}
\;.
\eeq
Then

\beq
\sigma_\rvb = \$(\rho_\rva)=
\sum_\mu K_\mu \rho_\rva K^\dagger_\mu
\;
\eeq
and

\beq
\sigma_{\rv{\mu}} = \$'(\rho_{\rv{\mu}})=
\sum_b L_b \rho_{\rv{\mu}} L^\dagger_b
\;.
\eeq

We've already shown how
density matrices
$\rho_\rva$ and $\sigma_\rvb=\$(\rho_\rva)$
can be represented by QB nets (see
Eqs.(\ref{eq-net-rhoa}) and (\ref{eq-net-sigb})).
Likewise, density
matrices $\rho_\mu$
and $\sigma_\mu=\$'(\rho_\mu)$
can be represented by
QB nets as follows:

\beq
\rho_{\rv{\mu}}= \tr_\rvj
\sandb{
\entrymodifiers={++[o][F-]}
\xymatrix{\rv{\mu}&\rvj\ar[l]}
}
\sandb{\hc}
\;
\eeq

\beq
\sigma_\rv{\mu}= \tr_{\rvb,\rvj}
\sandb{
\entrymodifiers={+++[o][F-]}
\xymatrix{
\cancel{\rvb}&*{}&\cancel{\rv{A}}\ar[dl]^<{0}&*{}\\
*{}&(\rvb,\rv{\mu})
\ar[ul]_>>{\delta}
\ar[dl]^>>{\delta}
&*{}&*{}\\
\cancel{\rv{\mu}}&*{}&\cancel{\rv{\mu}}\ar[ul]&\rvj\ar[l]
}
}
\sandb{\hc}
\;
\eeq

\end{document}